\documentclass[12pt,a4paper]{article}
\begin{document}
\sloppy
\title{{\bf Classical and Quantum Strings in plane waves, shock waves and space-time singularities : synthesis and new results}}
\author{Norma G. SANCHEZ \\
Observatoire de Paris, LERMA\\61, avenue de l'Observatoire\\75014 Paris, 
FRANCE\\Norma.Sanchez@obspm.fr}
\date{ }
\maketitle

\begin{center}
{\bf Abstract : }  
\end{center}
Key issues and essential features of classical and quantum strings in 
gravitational plane waves, shock waves and spacetime singularities are 
synthetically understood. This includes the string mass and mode number 
excitations, energy-momentum tensor, scattering amplitudes, vacuum 
polarization and wave-string polarization effect. The role of the real 
pole singularities characteristic of the tree level string spectrum 
(real mass resonances) and that of the spacetime singularities is clearly 
exhibited. This throws light on the issue of singularities in string 
theory which can be thus classified and fully physically characterized in two 
different sets : {\it strong} singularities (poles of order $\ge 2$, 
and black holes) where the string motion is {\it collective} and non 
oscillating in time, outgoing states and scattering sector do not appear, 
the string {\it does not} cross the singularities, and {\it weak} singularities (poles of order $<$ 2, (Dirac $\delta $ 
belongs to this class) and conic/orbifold singularities) where the whole 
string 
motion is oscillatory in time, outgoing and scattering states exist, and 
the string {\it crosses} the singularities. \\
Common features of strings in singular wave backgrounds and in inflationary backgrounds are explicitely exhibited. \\
The string dynamics and the scattering/excitation through the singularities 
(whatever their kind : strong or weak) is fully physically 
consistent and meaningful.\\ \\
\begin{center}
{\bf CONTENTS}
\end{center}
{\bf 1} - Introduction \\ \\
{\bf 2} - Understanding and Results \\ \\
{\bf 3} - Sourceless shock waves \\ \\
{\bf 4} - String Energy Momentum Tensor \\ \\
{\bf 5} - String Mass and Mode Number in Singular Plane-Wave Backgrounds \\ \\
{\bf 6} - Polarized plane waves and String Polarization \\ \\
{\bf 7} - Generic features. Common features with strings in inflationary 
backgrounds \\ \\
{\bf 8} - Generic wave profiles. The non-linear in-out unitary transformation   \\ \\
{\bf 9} - Clarification \\ \\
{\bf 10} - Conclusions and Classification 
  
\section{Introduction}
Classical and quantum strings propagating in gravitational plane waves, shock 
waves and space-time singularities and its classification, were investigated 
10 years ago by de Vega and Sanchez refs.\cite{vs1} - \cite{vs6} and by de Vega, Ramon Medrano and 
Sanchez refs.\cite{vms1} - \cite{vms3}, including the exact computation of the string mass spectrum, 
number operators, scattering amplitudes, string energy-momentum tensor,  
string vacuum polarization and string/wave polarization effects.\\ These 
results are relevant 
to answer questions as ``Singularities in String Theory'', and the 
boson-fermion and fermion-boson transmutations induced by superstring 
backgrounds.\\ The 
subject has subtle points and subtle physical interpretation which were 
overlocked in refs.\cite{hs1}, \cite{hs2}.\\ A lot attention has been payed recently to the subject \cite{prt1}, \cite{mt1}.\\
In this paper we provide a synthetic understanding to the problem and throw 
light on the issue of string singularities and its classification. 
\section{Understanding and Results}
We consider [1] strings propagating in gravitational plane-wave spacetimes 
described by the metric
\begin{equation}
dS^2 = F(U,X,Y)dU^2 - dU dV + dX^j dX^j,
\end{equation}
where
\begin{equation}
 F(U,X,Y) = W(U)(X^2-Y^2)\quad and \quad 2 \le j \le D-3. 
\end{equation}
$W(U)$ is an 
arbitrary function; $U$ and $V$ are light-cone variables. (These are 
sourceless solutions of the Einstein equations, i.e., vacuum spacetime). The 
only nonvanishing components of the Riemann tensor are 
\begin{displaymath}
R_{UiVj} = \frac{1}{2} \partial _i \partial _j F(U,X,Y)
\end{displaymath}
i.e.
\begin{displaymath}
R_{UXVX} = - R_{UYVY} = W(U)
\end{displaymath}
Thus, there will be spacetime singularities if $W(U)$ is singular. \\ \\
We are interested in functions $W(U)$ nonzero in a bounded interval 
$-T < U < T$ and which have a singular behavior for $U \rightarrow 0$, 
such as 
\begin{equation}
W(U) \sim \frac{\alpha }{|U|^\beta }\quad \mathrm{for}\quad U \rightarrow 0, 
\end{equation}
$\alpha $ and $\beta $ being positive contants. (A change of sign in 
$\alpha $ is just equivalent to exchanging $X$ with $Y$.) \\ \\ 
The essential features in this problem depend on the function W(U) (and not 
on the (X,Y) dependence). Moreover, our results hold for any function 
F(U,X,Y) = W(U) F(X,Y) (see below). The form of F(X,Y) determines whether or 
not the problem is exactly solvable but it is enough to fix the asymptotic 
behaviour of F(X,Y) to solve (asymptotically) the dynamics, to determine the 
behaviour near the space time singularities, and to determine whether 
$\langle M^2 \rangle $ 
and $\langle N \rangle $ will be (or will be not) finite. \\ \\
The string equations in the class of backgrounds eqs (1), (2) are linear 
and exactly 
solvable. In the light-cone gauge $U = \alpha' p \tau $ and after Fourier 
expansion in the world-sheet coordinate $\sigma $, the Fourier components 
$X_n (\tau )$ and $Y_n (\tau )$ satisfy a one-dimensional 
Schr$\mathrm{ \ddot{o}} $dinger-type equation but with $\tau $ playing the 
role of the spatial coordinate and $p^2 W(\alpha' p \tau )$ as the potential 
[2]. (Here $p$ stands for the $U$ component of the string momentum). \\ \\
The propagation of the string when it approaches the singularity at 
$U = 0 $ depends on the singularity exponent $\beta $ . We 
find different behaviors depending 
on whether $\beta < 2 $ or $\beta \ge 2 $ : \\ \\
(i) For $\beta < 2 $, the string coordinates $X$ and $Y$ are regular 
everywhere; 
that is, the string propagates {\it smoothly} through the gravitational 
singularity $U = 0 $. \\ \\
(ii) For a strong enough singularity ($\beta \ge  2 $), the string goes off to 
$X = \infty $ grazing the singularity plane $U = 0 $. This means that the 
string {\it does not go across} the gravitational wave; that is, the 
string 
cannot reach the $U > 0 $ region. For particular initial configurations, the 
string remains trapped at the point $X = Y = 0 $ in the gravitational-wave 
singularity $U = 0 $. \\ \\
(iii) The case in which $\beta = 2 $ and then 
$W(U) = \alpha /U^2 $ for all $|U| < T $ is explicitly solved in terms of 
Bessel functions. \\ \\
The string propagation in these singular spacetimes has common features with 
the fall of a point particle into a singular attractive potential 
$-\alpha / x^{\beta} $. In both cases, the falling takes place when 
$\beta \ge 2 $. \\ The behavior in $\tau $ of the string coordinates 
$X^A (\sigma , \tau )$ is analogous to the behavior of the 
Schr$\mathrm{ \ddot{o}}$dinger wave function $\Psi (x)$ of a point particle. 
{\it However, the physical content is different.} \\
The string coordinates $X^A (\sigma , \tau )$ are dynamical variables and 
not wave functions. Moreover, our analysis also holds for the quantum 
propagation of the string : the behavior in $\tau $ is the same as in the 
classical evolution with the coefficients being quantum operators. At the 
classical as well as at the quantum level, the string propagates or does 
not propagate through the gravitational wave depending on whether 
$\beta < 2 $ or $\beta \ge 2 $, respectively. In other words, the tunnel 
effect {\it does not take place} in this string problem. \\
\\ It must be noticed that for $\tau \rightarrow 0^- $, i.e., $U \rightarrow 
0^- $, the behavior of the string solutions is {\it 
non oscillatory} in 
$\tau $, whereas for $\tau \rightarrow \infty $, the string oscillates. This 
new type of behavior in $\tau $ is analogous to that found for strings in 
cosmological inflationary backgrounds. \\ \\
For $\beta = 2 $, we express the coefficients ($B^A _n $) characterizing the 
solution for $\tau \rightarrow 0 $ in terms of the oscillator operators for 
$\tau \rightarrow \infty $. We label with the indices $<$ and $>$ the 
operators in the region $U < - T $ and $U > T $, respectively (i.e., before 
and after the collision with the singularity plane $U = 0 $). \\ \\
For $\beta < 2$ we compute the total mass squared $\langle M^2 _> \rangle $ and the total 
number of modes $\langle N_> \rangle $ after the string propagates through 
the singularity plane $U = 0 $ and reaches the flat space-time region 
$U > T $. {\it This has a meaning only for} $\beta < 2 $. \\ \\
For $\beta \ge 2 $, the string {\it does not reach} the $U > 0 $ region 
and hence 
there are no outgoing operators ($>$). In particular no mass squared $M^2_>$ 
and total number $N_>$ operator can be defined for $\beta \ge 2 $. \\ \\
For $\beta < 2 $, $\langle M^2_> \rangle $ and $\langle N_> \rangle $ are 
given by [1].
\begin{equation}
\langle M^2_> \rangle = m^2_0 + \frac{2}{\alpha' } \quad \sum^\infty _{n=1} 
n(\left| B^x_n \right| ^2 + \left| B^y_n \right| ^2 ), 
\end{equation}
\begin{equation}
 \langle N_> \rangle = 2\quad \sum^\infty _{n=1}
(\left| B^x_n \right| ^2 + \left| B^y_n \right| ^2 ), 
\end{equation} 
where
\begin{equation}
B^x_n = -B^y_n \sim \left[ \frac{2n}{\alpha' p} \right]^{\beta - 2} 
\mathrm{for}\quad n \rightarrow \infty 
\end{equation}
($| 0_<  \rangle $ stands for the ingoing ground state, so that 
$\alpha_{n<} | 0_< \rangle = 0$ for all $n$). \\ \\
Therefore, $\langle M^2_> \rangle $ is {\it finite} for $\beta < 1 $ but 
{\it diverges} for $1 \le \beta < 2 $. $\langle N_> \rangle $ is finite 
for $\beta < \frac{3}{2}$ and diverges for $\frac{3}{2} \le \beta < 2 $. \\ \\
We analyze below the origin and physical meaning of these infinities. \\
Notice that the exponent $\beta $ in Eq.(6) is just minus the 
scaling dimension of $W(U)$ for small $U$. 
\section{Sourceless shock waves}
The sourceless shock-wave case 
\begin{displaymath}
W(U) = \alpha \delta (U),
\end{displaymath}
is particularly useful in order to understand the physical origin of the 
divergences in $\langle M^2_> \rangle $ when $1 \le \beta < 2 $ (and in 
$\langle N_> \rangle $ when $\frac{3}{2} \le \beta < 2$). \\ \\
For a sourceless skock wave with a metric function 
\begin{equation}
F(X,Y) = \alpha \delta (U)(X^2 -Y^2)
\end{equation}
In this $\delta $-function case, the $B_n$ coefficients are easy to compute 
exactly with the result 
\begin{displaymath}
B_n^x =- B_n^y = \frac{\alpha p \alpha' }{2in}
\end{displaymath}
Comparing the $B_n$ cofficients Eqs.(6), (8), we see that the large-$n$ 
behavior of 
$B_n$ for 
$W(U) = \alpha \delta (U)$, is the same as that for $W(U) = \alpha |U|^{-\beta }$ with $\beta = 1$. This is related to the fact that both functions $W(U)$ 
have the same scaling dimension. \\ \\
In this case the string propagation is formally like a Schr$\mathrm{\ddot{o}}$dinger equation with a Dirac $\delta $ potential : the string passes across the 
singularity at $U = 0 $ and tunnel effect is present.\\
This is a weak singularity, the behaviour of the string is oscillatory in 
time, and in, out scattering states can be defined. \\ 
The string 
scattering in this sourceless shock wave is very similar to the string 
scattering by a shock wave with a nonzero source density [2, 3, 6 ].\\ 
We also compute 
$\langle M^2_> \rangle $ and $\langle N_> \rangle $ for a metric function 
\begin{equation}
F(U,X,Y) = \alpha \delta (U)(X^2 - Y^2)\theta (\rho^2_0 - X^2 + Y^2), 
\end{equation}
where $\theta $ is the step function and $\rho _0$ gives the transverse size 
of the shock wave front. This $F$ belongs to the shock-wave class with a 
density source we have treated in refs.[2, 3, 6].\\ We find that 
$\langle M^2_> \rangle $ 
is {\it finite as long as $\rho _0$ is finite}. This divergence in 
$\langle M^2_> \rangle $ is due to the infinite transverse extent of the 
wave front and {\bf not} to the short-distance singularity $\delta (U)$ at 
$U = 0 $.\\ 
The gravitational forces in the transverse directions $X, Y$ 
transfer to the string a finite amount of energy when the transverse size of 
the shock wave ($\rho _0$) is finite. When $\rho _0 = \infty $, the energy transferred by 
the shock wave to the string produce large elongation amplitudes in the $X$ 
and $Y$ directions which are responsible for the divergence of 
$\langle M^2_> \rangle $.\\ \\ More generally, for a string propagating in a 
shock-wave spacetime with the generic profile
\begin{equation}
F(U, X, Y) = \delta (U) f (X, Y)
\end{equation}
we have found the {\bf exact} expression of $\langle M^2_> \rangle $ and 
$\langle N_> \rangle $ [6]. If $f(X, Y)$ is nonzero at $X = Y = \infty $, 
then $\langle M^2_> \rangle = \infty $.The divergence of ($M^2_>$) is due to 
the nonvanishing of 
$f(X, Y)$ at large distances and{ \bf not} to the singularity at 
U = 0 of the gravitational wave. When $f(X, Y)$ has an infinite range, the 
gravitational forces in the $X, Y$ directions have the possibility to 
transfer an infinite amount of energy to the string modes. 
\section{String Energy Momentum Tensor}
We also compute the string energy-momentum tensor $\mathcal{T} ^{AB}$. It is convenient to define 
it in the present context as an integral over a spatial volume completely 
enclosing the string at a time $X^0$, as we have proposed in Ref. 6. \\ \\
For $all \quad \beta >0$, we find that the $\mathcal{T} ^{AB}$ components 
can be grouped according to their $\tau \rightarrow 0$ behavior into four 
sets : \\ (i) $\mathcal{T} ^{VV}$ diverges for $\beta > 0$.\\ (ii) 
$\mathcal{T} ^{VX}, \quad \mathcal{T} ^{VY}$ diverge for $\beta > 1$ and tend 
to a finite constant for $\beta < 1$.\\ (iii) 
$\mathcal{T} ^{UV} \mathcal{T} ^Vj (3 \le j \le D - 1)$ and 
$\mathcal{T} ^{XX}$ are finite and nonzero. $\mathcal{T} ^{XX}$ vanishes for 
$\beta < 2$.\\ (iv) The other components vanish. \\ \\
The explicit expressions were obtained in ref. [1]. The energy density 
$\mathcal{T} ^{0 0} $, the energy flux in the propagation direction of the 
gravitational wave $\mathcal{T} ^{01}$, as well as the stress in this 
direction ($\mathcal{T} ^{11}$), strongly diverge for $\tau \rightarrow 0$. \\ 
For $\beta \le 2$, they diverge as a negative power of $\tau $. For $\beta > 2$, the divergence is exponential. \\ \\
We compute the ground-state expectation values in the illustrative case 
$\beta = 2$. The expectation values of {\bf all} the (operator) 
coefficients of 
the powers of $\tau $ turn out to be {\bf finite}. \\ \\
It can be noticed that the spatial proper length of the string grows 
indefinitely for $\tau \rightarrow 0$ when the string approaches the 
singularity plane. This phenomenon is analogous to that found for strings 
in cosmological inflationary backgrounds. \\ \\
In conclusion, for $\beta \ge 2$ and $\beta < 1$ the propagation of classical 
and quantum strings through the singular spacetimes Eqs. (1)-(3) is 
{\bf physically meaningful} and has the physical features described 
above.\\ \\ 
For $1 \le \beta < 2$, we find that the 
expectation value $\langle M^2_> \rangle $ diverges. However, this divergence 
is due to the infinite transverse extent of the wave front and {\bf not} to 
the 
short-distance singularity of $W(U)$ at $U = 0$. In support of this analysis, 
we found that the large-{\it n} behavior of the transmission coefficients 
(determining whether $\langle M^2_> \rangle$ is finite or infinite), is 
{\bf the 
same} for $W(U) = \alpha /U $ as for $W(U) = \alpha \delta (U)$.\\ Therefore,  the 
divergences of $\langle M^2_> \rangle $ in shock-wave spacetimes and in 
singular plane waves, have the same origin : the infinite transverse extension 
of the wave front.
\section{String Mass and Mode Number in Singular Plane-Wave 
Backgrounds}
As we have shown in ref.[1], the string performs strong oscillations in $X$ 
and 
$Y$ when it enters into the gravitational wave. These oscillations are 
stronger as the singularity at $\tau = 0$ is stronger. \\ \\
For $\beta \ge 2$, the string {\it does not cross} the $U = 0$ 
singularity and it 
escapes to $X = \infty $ grazing the singularity plane $U = 0$. \\ \\
For  $\beta < 2$, the string {\it crosses} the gravitational wave and 
reaches the 
flat space-time region $U > T $.\\ The expectation value of the string 
mass-square operator after the propagation through the gravitational wave has 
been computed in ref.[1]. \\ \\
The convergence of the series depends on the behaviour of Bogoliubov 
coefficients B$_n$ for large n. In this regime they can be computed in the 
Born approximation, with the result 
\begin{equation}
\mathrm{B_n} \sim -2i\alpha \left( \frac{2n}{\alpha' p}\right) ^{\beta - 2} 
C_{\beta }\quad \mathrm{for}\quad n \rightarrow \infty \quad 0 < \beta < 1 
\end{equation}
Therefore
\begin{displaymath}
n|B_n|^2 \sim n^{2\beta - 3}
\end{displaymath}
Hence,
\begin{equation}
\langle M^2_> \rangle = \left\{ \begin{array}{r@{\quad \quad}l}
finite\quad for\quad \beta < 1 \\divergent\quad for\quad 1\le \beta < 2 
\end{array} \right.
\end{equation} 
\\Recall that, when $\beta \ge 2 $, the string does not reach the region 
$U > 0$ 
and hence, there are no $>$ operators in the $\beta \ge 2$ case. \\ \\
For the expectation value of the number operator $\langle N_> \rangle $ we 
have 
\begin{equation}
\langle N_> \rangle = \left\{ \begin{array}{r@{\quad \quad}l}
finite\quad for\quad \beta < \frac{3}{2}, \\divergent\quad for\quad 
\frac{3}{2} \le \beta < 2.  \end{array} \right.
\end{equation}

The divergence appearing in $\langle M^2_> \rangle $ for 
$1\le \beta < 2$ is due to the infinite transverse extent of the wave front 
and {\bf not} to the short-distance singularity of $W(U)$ at $U = 0$. \\ \\
The gravitational forces in the transverse directions 
($X, Y$) transfer to the string a finite amount of energy when the size of 
the shock-wave front ($p_0$) is finite. The strong gravitational forces 
impart in the transverse directions ($X, Y$) large elongation amplitudes 
which are responsible for the divergent $\langle M^2_> \rangle $ when 
the size of the wave front is infinite. \\ \\
The divergent $\langle M^2_> \rangle $ for $W(U) = \alpha |U|^{-\beta }$ when 
$1 \le \beta < 2$ have the same explanation. (And similarly, those of 
$\langle N_> \rangle $ when $\frac{3}{2} \le \beta < 2$). \\ \\
More generally, for a string propagating in a wave type spacetime with 
generic profile
\begin{displaymath}
F(U, X, Y) = W (U) f(X, Y),
\end{displaymath}
if $f(X, Y)$ does not vanish at $X = Y = \infty $, then 
$\langle M^2_> \rangle = \infty $.
\section{Polarized plane waves and String Polarization}
The above properties can be generalized to singular gravitationaal waves 
with two arbitrary singular profile functions $W_1(U) = \alpha _1/|U|^{\beta _1}$, 
$W_2(U)  = \alpha _2/|U|^{\beta _2}$. \\ \\
The dynamics is exactly and explicitely solvable [7]. The string time 
evolution is fully determined by the background geometry, whereas the overall $\sigma $-dependence is fixed by the initial string state. \\ \\
The proper length stretches infinitely at the singularities when 
$\beta _1 \ge 2$ and / or $\beta _2 \ge 2$ (strong singularities)\\ \\
When $\beta _1 \ge 2$ and / or $\beta _2 \ge 2$, the string does not cross 
the singularity (U = 0) but goes off to infinity in a given direction 
$\alpha $ which depends on the polarization of the gravitational wave. The 
string escape situation is the following : \\
(i) for $\beta _1 > \beta _2$ (and $\beta _1 \ge 2$), then the angle 
$\alpha = 0$ and the string goes off to infinity in the X-direction. In this 
case, the singularity of $W_1$ dominates over that of $W_2$ and we recover 
the previous situation of ref. \cite{vs1}.\\ \\
(ii) for $\beta _2 > \beta _1$ (and $\beta _2 \ge 2$), then 
\begin{equation}
\alpha = (\pi/4)\quad \mathrm{sgn} \quad \alpha _2,
\end{equation}
and \\ \\
(iii) for $\beta _1 = \beta _2 \ge 2$, then 
\begin{displaymath}
\tan \alpha = \frac{\alpha _2}{\alpha _1 + \sqrt{\alpha ^2_1 + \alpha ^2_2}}
\end{displaymath}
that is 
\begin{displaymath}
\tan 2 \alpha = \alpha _2/\alpha _1.
\end{displaymath}
If $\alpha _1  > 0 \quad (\alpha _1 < 0)$, the string escape directions are 
within the cone $|\alpha | < \pi /4 \quad (|\alpha - \pi /2| < \pi /4)$. \\
In addition to escaping to infinity, the string oscillates in the (X, Y) 
plane, perpendicularly to the escape direction, and with vanishing 
amplitude for $U \rightarrow 0$.\\
The string behaviour near the singularity expresses naturally in terms of the null variable $\hat{U}$. 
\begin{displaymath}
\hat{U}(\sigma , \tau ) = \left\{ \begin{array}{r@{\quad \quad}l}
\ln (-U) \quad \beta = 2\\
\frac{(-U)^{1-(\beta /2)}}{1 - (\beta /2)} \quad \beta > 2 \end{array} 
\right.
\end{displaymath}
For instance, the oscillatory modes in the (X, Y) plane are not 
harmonic in $U$ but in $\hat{U}$. The variable $\hat{U}$ is like the cosmic 
time of strings in cosmological backgrounds (in terms of which the string 
oscillates), whereas $U$ is like the conformal time. Here, for simplicity, 
$\beta _1 = \beta _2 = \beta $ but this case is actually generic. \\ \\
For $\beta _1 < 2$ and $\beta _2 < 2$ (weak singularities), the string 
passes smoothly through the space-time singularity and reaches the 
outgoing region, and, the string behaviour is oscillatory in time. In this 
case, outgoing operators make sense and can be explicitely related to the 
in-operators. \\ 
For the particles described by the quantum string states, this implies two types of effects : \\
(i) rotation of spin polarization in the (X, Y) plane, and \\
(ii) transmutation between different particles.\\ \\
The expectation values of the outgoing mass ($M^2_>$) operator and of the mode-number operator $N_>$, in the 
ingoing ground state $|O_< \rangle $ are different from the ingoing 
expectation values $M^2_<$ and $N_<$. This difference is due to the 
excitation of the string modes after crossing the spacetime singularity. 
In other words, the string state is not an eigenstate of $M^2_>$, but an 
infinity superposition of one-particle states with different masses. This 
is a consequence of the particle transmutation which allows particle masses 
different from the initial one ($m^2_0$). 
\section{Generic features. Common features with strings in inflationary 
backgrounds}
The string evolution near the spacetime singularity is a {\it collective 
motion} governed by the nature of the gravitational field. The state of the 
string fixes the overall $\sigma $-dependent coefficients whereas the 
$\tau $-dependence is fully determined by the spacetime geometry. In other 
words, the $\tau $-dependence is the same for all modes $n$. \\ 
In some directions, the string collective propagation turns to be an 
infinite motion (the escape direction), whereas in other directions, the 
motion is oscillatory, but with a fixed ($n$-independent) frequency. In fact, 
these features are not restricted to singular gravitational waves, but are 
{\it generic} to strings in strong gravitational fields . \\ \\
Moreover, it is interesting to compare the string behaviour for 
$\tau \rightarrow 0$ in the inflationary backgrounds for which we have ref.[12]
\begin{displaymath}
ds^2 = (dX^0)^2 - R^2(X^0)(dX^i)^2 
\end{displaymath}
with $R(\tau ) = -(H \tau )^{-\beta/2}$
\begin{displaymath}
X^0(\sigma , \tau )_{\tau \rightarrow 0} = \left\{ \begin{array}{l@ {\quad} r} 
\mathrm{const} \quad \quad \quad \quad \quad \quad \quad \beta < 2 \, 
\mathrm{(superinflationary)} \\
-H^{-1} \ln [H \tau L(\sigma )] \quad \quad \beta  = 2 \, \mathrm{(de 
\, Sitter,  inflationary)} \\
-\frac{(\tau L(\sigma ))^{1-(\beta /2)}}{1-(\beta /2)} \quad \quad \quad 
\beta > 2  \mathrm{(power \, type, \, inflationary)}
\end{array} \right.
\end{displaymath}.
\begin{displaymath}
X^i(\sigma , \tau )_{\tau \rightarrow 0} = \left\{ \begin{array}{l@ {\quad} r}
A^i(\sigma ) + D^i(\sigma )\frac{\tau ^2}{2} + \tau ^{1+\beta }F^i_{(\sigma)} 
\quad \quad \quad \beta \neq 2 \\
A^i(\sigma ) +\tilde{D}^i(\sigma )\frac{\tau ^2}{2} + \frac{\tau ^2}{2}
\ln \tau \tilde{F}^i_{(\sigma )} \quad \quad \beta = 2.
\end{array} \right. 
\end{displaymath}
The $A^i(\sigma )$ are arbitrary functions of $\sigma,$ $D^i(\sigma ),$ 
$\tilde{D}^i(\sigma ),$ $F^i(\sigma),$ and $\tilde{F}^i(\sigma )$ are fixed by the constraints. Here $X^0$ is the cosmic time, while $\eta \approx \tau $ 
is the conformal time.\\ For the singular plane waves we have 
\begin{displaymath}
\hat{U}(\sigma , \tau )_{\tau \rightarrow 0} = 
\left\{ \begin{array}{l@ {\quad} r}
\mathrm{const} \quad \quad \quad \quad \quad \beta < 2 \\
\ln (-U) \quad \quad \quad \quad \beta = 2 \\
\frac{(-U)^{1 - (\beta / 2)}}{1 - (\beta / 2)} \quad \quad \quad \beta > 2.
\end{array} \right.
\end{displaymath}
The behaviour of the string time coordinates are {\it regular} and 
{\it non-oscillating} for $\tau \rightarrow 0$ ('frozen'), while for 
singular 
plane waves, in addition, one of the transverse coordinates is non-oscillating and singular for $\tau \rightarrow 0$. \\ \\
The above behaviours for $X^0$ and $U$ are characteristic of strings in strong gravitational fields. Notice that the inflationary backgrounds are non-
singular, whereas the plane waves (equations (1)-(2)) are singular 
spacetimes. This is connected to the fact the string coordinates $X$, $Y$ 
exhibit divergences in these plane-wave spacetimes. 
\section{Generic wave profiles. The non-linear in-out unitary transformation}
In the case of plane or shock waves with profile functions ($X^2 \pm Y^2$), 
the string equations became {\it exactly linear}, and the transformation 
between ingoing ($\langle $) and outgoing operators ($\rangle $) is 
{\it linear}, ie a Bogoliubov transformation. \\ \\
For generalized profile functions of plane waves and shock waves, the string 
equations are non linear and the ingoing-outgoing unitary transformation 
describing the scattering/excitation of the string is {\bf non-linear}. 
The linear (Bogoliubov) approximation holds only for large impact parameters. 
One must use the {\bf exact} in-out transformation to include {\bf all} 
impact parameters. In ref \cite{vs6} we succeeded in computing the 
{\bf exact} transformation. This throws light on both computations and 
interpretation, and in particular on the r\^{o}le played by the spacetime 
background. We do all the treatement for any function $f(X^i)$, i.e. for 
any wave profile. We express the nonlinear transformation between 
ingoing and outgoing zero modes and oscillators in terms of a hermitean 
operator
\begin{displaymath}
G = \frac{p_U}{8\pi } \int^{2\pi }_0 \: d\sigma \, \int d{\bf p}^{D-2} \, : 
\, e^{{\bf ip-x}(\sigma ,\tau =0)}\, : \, \varphi ({\bf p}), 
\end{displaymath}
where : : stands for normal ordering with respect to the ingoing state 
$|0_< \rangle $ and $\varphi$({\bf p}) is the Fourier transform of the 
wave profile. \\ \\
This operator acts on the Fock space operators and Fock space states and 
it is suitable to express the relevant expectation values after the 
collision of the string with the wave background.\\
The ingoing-outgoing ground-state transition amplitude $\langle 0_<|0_>\rangle 
$ expresses as 
\begin{displaymath}
\sum _n \frac{i^n}{n!} \langle 0_<|G^n|0_< \rangle \,.
\end{displaymath}
We interpret these terms as a $n$-leg scalar amplitude with vertex operators 
inserted at $\tau $ = 0 (a line of pinchs at the intersection of the world 
sheet with the wave), and find an integral representation for these 
quantities. \\ \\
The integrands possess equally spaced real pole singularities typical of 
string models in flat space-time. The presence and structure of these 
poles is not at all related to the structure of the space-time geometry 
(which may or may not be singular). We give a sense to these integrals by 
taking the principal value prescription, yielding for 
$\langle 0_< |G^n|0_< \rangle $ a well defined {\it finite} and real 
result. \\ \\
The exact expressions for the total outgoing mode operator $\langle N_>
\rangle $ and mass square $\langle M^2_> \rangle $ operator in the ingoing 
ground state $|0_< \rangle $ can be thus computed. We find that the 
contribution from each $n$-level, that is $M^2_n $ for $n \rightarrow \infty $, decreases like 
\begin{equation}
\alpha ' \tilde p^2 n^{-1} (2\alpha ' / \pi \log n)^{1-D/2}
\end{equation}
The large $n$ behaviour of $\langle M^2_n \rangle $ depends on the density 
matter $\tilde{\rho } $ only through its total energy 
$\tilde{\rho } = \int d^{D-2} {\bf X} \, \tilde{\rho }({\bf X})$. \\ \\
All the shock wave geometries exhibit the same large $n$ behaviour of 
$M^2_n $, irrespective of the structure of the localized source. \\
The sum over $n$ in both $\langle N_> \rangle $, and 
$\langle M^2_> \rangle = (D-2)/(12\alpha ') + \sum^{\infty }_{n-1} 
\langle M^2_N \rangle $ converges and the total values are exactly computed. \\ \\
The contribution of the excited states is suppressed by a factor 
$\langle 0_<|0_< \rangle = (L/2\pi )^{D-2}$ with respect to the direct 
(initial state) contribution $\mu ^2 = -(D-2)/(12\alpha ')$, $L$ being a 
typical large box size. \\ \\  
We find for $\langle N_> \rangle $ and $\langle M^2_> \rangle $ integral 
representations which exhibit a similar structure to $\langle G^2 \rangle $, 
that is, the integrands factorizes into two pieces  : $|\phi ({\bf p})|^2 $ 
which characterizes the wave geometry and the function 
\begin{displaymath}
\mathrm{tg}(\alpha '\pi {\bf p}^2)\Gamma (\alpha '{\bf p}^2)/\Gamma 
\left( \frac{1}{2} + \alpha '{\bf p}^2 \right) \, ,
\end{displaymath}
which depends only on the string. These integrands possess real singularities 
(poles) like the tree level string spectrum. 
This is exactly what happens in the tree amplitudes of string models. Like the tree level string spectrum, these poles correspond to equally spaced real resonances in the mass 
spectrum. As is usually expected, loop corrections provide a width to these 
resonances and will therefore shift the poles away from the integration path, leading to finite results. The physical interpretation of such poles is that 
they correspond to all higher string states which become excited after the collision through the wave singularities.\\
In conclusion, singularities in string theory are fully, mathematically  and 
{\bf physically consistent}.  
\section{Clarification} 
Some aspects of strings in these gravitational-wave backgrounds have been 
studied by the authors of refs.\cite{hs1}, \cite{hs2}. 
However, this problem has subtle points which were overlooked in refs.
\cite{hs1}, \cite{hs2}.\\ The 
analysis done in refs. \cite{hs1}, \cite{hs2} by analogy with the Schr$\mathrm{\ddot{o}}$dinger 
equation is not careful enough. The mass and number operators are expressed 
in terms of the transmission coefficient $B_n$. The cases in refs. \cite{hs1}, \cite{hs2}  in which $B_n = \infty $ mean that there is no transmission to 
the region $U > 0$, 
and therefore that there is no outgoing mass nor outgoing number operator 
(since there is no string) in 
that region. This is the situation of string falling to $U = 0$ for $\beta \ge 2$, the string {\it does not cross} the singularity, the string scapes 
to infinity {\it grazing} the singularity plane U = 0.\\ 
Therefore,the outgoing ($>$) operators as $M^2_> $ and $N_> $ make sense only 
for $\beta < 2$ and any statement about $M^2_> $ and $N_> $ for $\beta \ge 
2$ is {\bf meaningless}, (since there is no string in the $>$ region for 
$\beta \ge 2 $). \\ \\
In the cases in which $\langle M^2_> \rangle $ is divergent, such infinity is 
{\bf not} related at all to the spacetime singularity at $U = 0$. This happens for 
$\beta < 2$ when the transverse size (i.e., perpendicular to the propagation 
direction) of the gravitational wave front is infinity. Then, the 
gravitational wave carries an infinite energy which transfers to the string 
according to the behavior of $W(U)$. This may lead to a finite or infinite 
value for $\langle M^2_> \rangle $ as we have seen above.\\ \\
In conclusion, the propagation of classical and quantum strings through the 
singular space-times is {\bf physically meaningful} and follows the 
evolution, properties and classification summarized here.
\section{Conclusions and Classification}
{\bf (i)} As is known, in the context of point particle QFT, vacuum polarization 
effects do {\bf not} arise in plane wave and shock wave backgrounds 
since ingoing $<$\, and outgoing \,$>$ operators {\bf do not} get mixed in 
this context. Therefore no particle creation effects takes place for point 
particle field theories in these geometries. On the contrary, particle 
transmutations as well as vacuum polarization effects on the energy-
momentum tensor {\bf do appear} for strings in these wave space-times. 
These effects can be traced back to the mixing of creation and 
annihilation $<$\, and\, $>$ string oscillators. \\ \\ 
{\bf (ii)} Pole-type  curvature singularities with $\beta < 2$, $\delta (U)$ 
singularities conical and orbifold singularities are {\bf weak} singularities : the string motion is oscillatory in time, string crosses smoothly the 
singularities, ingoing and outgoing states can be defined and so all the 
scattering sector. Mass, number and energy-momentum of the string are well 
defined and {\bf finite}. The string oscillators get excited in the 
scattering and crossing the singularities. The string proper length is 
finite. The detailed classical and quantum string dynamics in conical space 
times with general deficit angles
\begin{displaymath}
\delta \phi = 2 \pi (1 - \alpha ) = 8\pi G\mu 
\end{displaymath}
\begin{displaymath}
0 \le  \phi < 2\pi \alpha ,
\end{displaymath}
was treated in refs \cite{vs7}, \cite{vms4}. In the simpler case of orbifolds, which is a particular case 
of the above for deficit angles $2\pi (1 - 1/N)$, the scattering is trivial. \\
\\ {\bf (iii)} Pole-type singularities with $\beta \ge 2$ and black holes (r = 0) are 
{\bf strong} singularities; string does not pass across the singularity 
(eventually it can gets trapped); outgoing operators can not be defined. 
Ingoing mass, number and energy momentum tensor of the string are well 
defined and {\bf finite}. \\
The proper string length stretches indefinitely. \\ \\
{\bf (iv)} The features of the strong singularity case are {\bf generic} to 
strings in strong gravitational fields : the string evolution near the strong 
spacetime singularity is a {\bf collective motion} governed by the nature of the gravitational field. The state of the string fixes the overall 
$\sigma $-dependence. The time dependence is fully determined by the 
spacetime geometry, that is, the same for all modes $n$. In some directions, 
the string collective propagation is an infinite motion (the escape direction); in other directions, the motion is oscillatory, but with a fixed ($n$)-
independent frequency. \\
This happens too for strings in inflationary backgrounds (which are not 
singular). 
In the inflationary cases, all transverse spatial coordinates are 
{\bf non oscillating} in time, but they are {\bf regular} (``frozen'') 
for $\tau \rightarrow 0$, while for strong singular plane waves some of the 
transverse coordinates are singular. \\
For the string time coordinate the plane wave case $\beta < 2$ is like 
super inflation, the case $\beta = 2$ is like de Sitter inflation, and 
the case $\beta > 2$ is like power type inflation. \\ \\
{\bf (v)} Tests strings do propagate {\bf consistently} in singular space-times : 
Klein-Gordon equation (for a point particle) is ill defined, whereas the 
string equations are well behaved. This includes strings in plane wave 
backgrounds, shock wave  backgrounds (sourceless shock waves and of Aichelburg-Sexl 
type, and with generic profiles), as well as in the black hole geometries, where the string behaviour is 
{\bf regular} at the horizon and near the r = 0 sigularity refs.\cite{vs8}, 
\cite{vs9}, \cite{ls1}. That is, strings feel the space time singularities 
much less than point particles. \\
Furthermore, we would not be surprised by the presence of space-time 
singularities in string theory as long as one sticks to a geometry 
description using a metric tensor $G_{AB}(X)$ in spite of the fact it 
fulfills the exact or corrected string equations. We do not expect that a 
space-time description in terms of a Riemannian manifold with local 
coordinates $X^A$ will be meaningful at the Planck scale.

\end{document}